\documentclass[aps,twocolumn]{revtex4}
\usepackage{graphics}
\usepackage{epsfig}
\usepackage{dcolumn}
\usepackage{amsmath,mathptm}
\usepackage[usenames]{color}

\begin{document}

\title{Exact dynamics for fully connected nonlinear networks }

\author{G. P.  Tsironis}
\address{Department of Physics, University of Crete and Institute of Electronic Structure and
Laser, FORTH, P.O. Box 2208, Heraklion 71003, Crete, Greece.}




\begin{abstract}

We investigate  the dynamics of  the discrete nonlinear Schr\"{o}dinger  equation in fully connected networks.  For a localized initial condition  the exact solution shows the existence of two dynamical transitions as a function of the nonlinearity parameter, a hyperbolic and a trigonometric one.
In the latter the network behaves exactly as the corresponding linear one but with a renormalized frequency.
 
\end{abstract}

\maketitle


\section{Introduction}
Nonlinear dynamics in complex networks  incorporates competition of propagation, nonlinearity and
bond disorder and may find some applications in complex natural or man-made materials.  
For a given random network of
$N$ sites the limit of fully coupled lattice where each site is connected to every other one with the same strength 
plays an important role
since, for zero nonlinearity, introduces a high degree of degeneracy and thus localization~\cite{PT11}.  It is 
therefore interesting to probe the dynamics in this fully coupled limit when the network is nonlinear.  We use the Discrete Nonlinear Schr\"odinger (DNLS) equation for this study since the latter is a prototypical equation with a large number of applications in condensed matter physics, optics, Bose-Einstein condensation, etc.~\cite{ELS85,AK98,Led08}.

The DNLS equation in the fully coupled or Mean Field (MF) limit is given by:

\begin{eqnarray}
i\frac{d\psi_n}{dt}=\epsilon_n \psi_n + V\sum_{m \ne n}  \psi_{m}-\gamma |\psi_n|^{2} \psi_n
\label{dnls1}
\end{eqnarray}

where $\psi_n(t)$ is a complex amplitude,  $V$ denotes the constant overlap integral connecting any two sites,
$\epsilon_n$ a local site value
 and $\gamma$ is a parameter that
controls the strength of the nonlinear term.  When $t$ is time, the equation describes the dynamics of
interacting particles with the nonlinear term representing polaronic effects or interaction of
bosons in an optical lattice.  When $t$ is a space variable, the equation describes electric wave modes propagating in fibers that include nonlinearity.
In the present analysis we take $\epsilon_n =0$ while the DNLS norm equal to one, viz.
 $\sum|\psi_n|^2 = 1$ . For notational simplicity we  rescale time to $\tau = V t$ and $\gamma$ to
$\chi =\gamma /V$; we  note that for $\chi > 0$  we have the 
defocusing DNLS equation while for $\chi < 0$ we obtain the focusing case.

\section{Review of earlier literature}
The problem of the DNLS equation  in fully coupled lattices has been addressed in the past~\cite{ELS85,MT93,AK93a,AK93,M99}.
Eilbeck et al.~\cite{ELS85} analysed the stationary states of the problem for the general $N$-site case. Molina and Tsironis
~\cite{MT93} showed that in the case of $N=3$, i.e. a symmetric trimer, and for a localized initial condition,  a   transition occurs for $\chi =-6$ where the evolution  ceases to be periodic and the averaged
probability becomes equal in all three sites.  It was also pointed out that 
no selftrapping occurs for positive nonlinearity values. This dynamical behaviour
is analogous, but not identical, to the well known selftrapping transition of the dimer~\cite{KC86,TK88}. 
Andersen and Kenkre~\cite{AK93a} solved analytically the symmetric trimer case for localized as well as symmetrically delocalized initial
conditions. They wrote explicitly the time evolution at selftrating ($\chi =-6$), connected it to the stationary states, calculated the averaged occupation probability of the initially populated  site and showed that for a given value of the nonlinearity parameter extreme delocalization occurs. They also pointed out that the nature of selftrapping in the trimer was distinct from that in the dimer case.  Subsequently, Andersen and Kenkre~\cite{AK93} analysed the fully connected $N$-site
problem using similar methods for localized and partially localized initial conditions and showed that a "migratory" 
transition takes place in some cases. The latter is related to the fact that localization of the initially non-occupied
sites may occur for some parameter values. In ~\cite{AK93} a general expression in terms of the Weierstrass elliptic function
is given for the time evolution of the occupation probability as well as expressions for some special cases. Additionally the time-averaged occupation probability is analysed. Finally, Molina ~\cite{M99} used an extension of the method 
of  ref. ~\cite{MT93} and determined explicitly the values of the nonlinearity parameter at which there 
is a "transition" for arbitrary values of $N$.

From the previous review of the existing literature it is clear that the problem of nonlinearity induced localization in the fully coupled, MF  limit, has been analysed thoroughly and many of the features of the solution have been already discovered. However, we believe that  a complete dynamical picture has not emerged yet.  Thus we will follow the approach of ref.
~\cite{AK93}, re-derive the complete dynamical solution and use it to focus on the explicit dynamics and not just to the 
average occupation probability.

\section{Nonlinear Localization}

In the purely linear case  for $\chi =0$, Eq. (\ref{dnls1}) reduces to a linear eigenvalue problem that
has $N-1$ fold degenerate eigenvalues~\cite{PT11}. This high degree of degeneracy forces 
an initially  spatially localized  state to remain localized executing incomplete oscillations to the other, initially unoccupied,  sites. This linear localization at the initially excited site increases with the
system size.

\subsection{Nonlinear mean field limit}
 
  For the localized initial condition we observe that if we set initially all probability at site one, then due to the
symmetry of the lattice, the subsequent evolution of all sites except the initially populated one will be identical; as a 
result we can cast the original $N$-equation problem to a simpler one consisting of only two equations, viz.~\cite{MT93,AK93a,AK93,M99}
\begin{eqnarray}
i\dot{\psi_1}=(N-1) \psi_2 -\chi |\psi_1 |^2 \psi_1 \\
i\dot{\psi_2} =\psi_1 +(N-2) \psi_2  -\chi |\psi_2 |^2 \psi_2
\label{dnls2}
\end{eqnarray}   
with $|\psi_1 |^2 + (N-1) |\psi_2 |^2 =1$.  To obtain the set of equations \ref{dnls2} we have set
$\psi_2 = \psi_3 = ..=\psi_N$ at all times.  We proceed by introducing the variable $p(t) = |\psi_1|^2 -(N-1) |\psi_2|^2$
and cast the set of complex equations (\ref{dnls1}) first into a  second order equation that has the form of
Newton's law and subsequently into an equivalent potential problem for the localized initial condition $p(0)=1$~\cite{KC86,MT93,AK93a,AK93,M99}, viz.
\begin{widetext}

\begin{eqnarray}
{\dot{p}}^2 =U(p) \equiv a_0 p^4 +4 a_1 p^3 +6 a_2 p^2 +4 a_3 p + a_4 \\
a_0 =  -\frac{\chi^2 N^2}{16(N-1)^2}\\
a_1 =  -\frac{1}{48 (N-1)^2 } \left[ 3 \chi (N-2) N(-2+\chi +2 N) \right] \\ \\
a_2 = \frac{1}{48 (N-1)^2 } [ 32 \chi -8\chi^2 -56\chi N  \nonumber\\
 +4\chi^2 N -8 N^2 +28 \chi N^2 +\chi^2 N^2 +16 N^3 - 4\chi N^3 -8N^4 ]\\
a_3 = \frac{1}{16 (N-1)^2 } \left[ (N-2)(2N-2 +\chi )(8-4\chi -12 N +3 \chi N +4 N^2 )\right] \\
a_4 = \frac{1}{16 (N-1)^2 }[ 128-64\chi +16 \chi^2 -384 N+144\chi N -24\chi^2 N + \nonumber\\
400 N^2 -104 \chi N^2 + 9\chi^2 N^2 -160 N^3 +24 \chi N^3 +16 N^4 \\
\label{potential}
\end{eqnarray}
\end{widetext}
We now form the invariants $g_2$, $g_3$:
\begin{eqnarray}
g_2 = a_0 a_4 -4 a_1 a_3 +3 a_2^2\\
g_3 =a_0 a_2 a_4 +2 a_1 a_2 a_3 -a_2^3 -a_0 a_3^2 -a_1^2 a_4
\label{invariants}
\end{eqnarray}
The solution is given in terms of the Weierstrass elliptic function ${\cal{P}} (t , g_2 , g_3 )$ as follows
~\cite{AS65}:
\begin{eqnarray}
p(\tau )=1-\frac{24(N-1)}{12{\cal{P}} (\tau  , g_2 , g_3 ) +\chi^2 +2(N-2)\chi +N^2 }
\label{solution}
\end{eqnarray}
This solution depends only on the nonlinearity parameter $\chi$ and the number of sites $N$ and presents
the exact solution of the nonlinear DNLS problem in the MF limit for localized initial condition and arbitrary site number; it is equivalent to  the solution  in ref. ~\cite{AK93} for perfect initial localization on a single site.
The Weierstrass function is doubly periodic with periods $2\omega$ and $2\omega '$ where $\omega / \omega '$ is not real. 
The discriminant $\Delta = g_2^3 -27 g_3^2$ may be expressed as follows
\begin{widetext}
\begin{eqnarray}
\Delta (N) =-\frac{1}{16}N^2 \chi^2 g(N) \left( \chi + 2N \right) \\
g(N)= 2( N-2)\chi^3 +
2 \left( N-2\right) \left( 3N-8\right) \chi^2 +\left( 6 N^3 -53 N^2 +112 N -64 \right) \chi 
+2 (N-1) N^3
\label{discriminant}
\end{eqnarray}
\end{widetext}
When the discriminant becomes zero, the Weierstrass solution changes character signalling a change
in the physical behaviour of the system. The quantity $\Delta(N)$ has three real roots, denoted hereafter
as $\chi_0$, $\chi_c$ and $\chi_c '$ for $\chi_0 =0$, $g(\chi_c )=0$ and $\chi_c ' =-2N$ respectively.

The root for $\chi_0 =0$ corresponds to the linear MF problem resulting in $g_2 = N^4 /12$, $g_3 =N^6/216$, the  Weierstrass function 
becomes  a trigonometric function, viz. ${\cal{P}} (t, g_2 , g_3 )=-N^2/12 +N^2/4 \sin^2 \left[ (N t/2)\right] $  and the
solution for $p(\tau )$ is 
\begin{eqnarray}
p(\tau ) \equiv p_{lin}(\tau )= \frac{(N-2)^2 + 4(N-1) \cos (N \tau ) }{N^2}
\end{eqnarray}
a result that agrees with the linear solutions derived in ref. ~\cite{PT11}.  We note that in the linear case the 
system executes oscillations between $p=1$ and  $p_{lin,M}=\left[ (N-2)^2 - 4 (N-1)\right] /N^2$
while $p(\tau )$ oscillates around $<p_{lin} > =(1-2/N)^2$.  Only for $N=2$ we have complete oscillations in the $p$-range
$1,~-1$ while for all other geometries the linear system oscillations are incomplete, viz. there is always some 
occupation probability  at the originally populated site.  In the limit of very large $N$ there is extreme localization at the initially populated site.

In order to understand the role of $\chi_c$, $\chi_c '$ in the dynamical evolution, 
it is instructive to address first explicitly the regime
 for $N \leq 4$; although these cases have been studied  in the past, the
present exposition furnishes additional results and new intuition. 
\subsection{$2\leq N \leq 4$}
In this range we have  $\chi_c ' \leq \chi_c$ with the equality holding for $N=4$. 
\subsubsection{N=2}
In the dimer case  we have $\chi_c = 4$ and $\chi_c ' =-4$; when $\chi$ assumes one of these two critical values, the invariants become $g_2 =4/3$, $g_3 = -8/27$ , 
the frequencies $\omega = {\bf K} (1) =\infty $, $\omega ' = i{\bf K'} (1)=i\pi/2$ and the Weierstrass function becomes a hyperbolic function, viz. ${\cal{P}} (t, 4/3,-8/27 ) = 1/3 + 1/\sinh^2 t$ leading to $p(t)= sech (2t)$.
This situation corresponds to the well known  selftrapping transition for a localized initial condition where
the solution at the transition is expressed as hyperbolic secant~\cite{KC86}. We note that in the dimer case
$\chi_c =| \chi_c ' |$ and for $-\chi_c < \chi< \chi_c $  the particle executes complete oscillations.
\subsubsection{N=3}
In ref. ~\cite{MT93} a  transition was found occurring at $\chi_c ' =- 6$, while in refs. ~\cite{AK93a,AK93} the dynamical study was  performed. Analysis of the discriminant shows that in addition to the previously identified root $\chi_c ' =-2 N = -6$ there is also the
real root of $g(\chi_c )=0$  at
\begin{eqnarray}
\chi_c = -\left[ 1+\frac{131}{\left( 6614-774\sqrt{43}\right)^{1/3}}+\frac{\left( 3307-387\sqrt{43}\right)^{1/3}}{3^{2/3}}\right] 
\label{trimer-root}
\end{eqnarray}
or $\chi_c \approx -6.039904249$. 

For $\chi =\chi_c ' \equiv -6$ the invariants are $g_2 =3/4$, $g_3 = -1/8$, $\omega$ becomes infinite  and
the Weierstrass function turns into a hyperbolic function, viz. 
${\cal{P}} (t , 3/4 , -1/8 ) = 1/4 + 3/4 \left[  sinh (\sqrt{3/4}t ) \right] ^{-2} $.  The time evolution for $p(\tau )$
becomes~\cite{AK93a}
\begin{eqnarray}
p(\tau )=\frac{1}{3}\left[ \frac{3 - 4 sinh^2 (\sqrt{3/4}\tau  )}{1+4  sinh^2 (\sqrt{3/4}\tau  )}\right] 
\end{eqnarray}
At long times $p(\tau )_{lim \tau \rightarrow\infty } =-1/3 \approx -0.333$ corresponding to equal probability $1/3$ on all three sites and should be compared with   $p_{lin,M} =-7/9 \approx -0.777$ and
$<p_{lin}> =1/9 \approx .111$
When $\chi = \chi_c $  both invariants are positive and the
Weierstrass function turns into a simple trigonometric function.

We now describe the dynamics of the trimer as a function of the nonlinearity parameter $\chi$.    
For $\chi =0$,  $p(\tau )$ performs incomplete oscillations between
$p(0)=1$ and $p_{lin,M}=-7/9$.  Increasing  $\chi$ to positive values augments gradually the effect
of localization  and the particle stays almost completely at the initial
site for large values of the nonlinearity parameter.  For negative values of $\chi$, however, the behaviour
is markedly different; upon increasing $\chi$ in absolute value,  the particle becomes more detrapped and performs
oscillations beyond the $p_{lin,M}$ limit. This behaviour continues as we further increase $|\chi |$ and at $\chi = -4$
the particle executes  complete oscillations to the other sites.  Further increase
of nonlinearity reduces the amplitude of the oscillatory motion while the oscillation period increases. For $\chi = \chi_c'$ there is a transition to a hyperbolic, non oscillatory behaviour; the particle collapses from complete oscillations to complete equipartition of probability in the three sites (at infinite times). This is a remarkable transition since it localizes again the particle and all motion stops asymptotically~\cite{MT93,AK93a}.
As the value of $|\chi|$  increases further, the trapping tendency increases rapidly  and the particle becomes  
more trapped in the initial site.

As the trimer crosses the value $\chi_c $ the discriminant becomes zero again with positive invariants; this
means that the time evolution simplifies to one with  trigonometric evolution. Although this feature is not
exceptional in the trimer, it is quite interesting when it appears for $N>4$; we will comment
on this trigonometric transition  below.

The transition at $\chi_c' $ is  a  selftrapping transition in the sense that it  forces  
probability equipartition on all three sites, similar to the corresponding behaviour in the dimer. It is remarkable
that nonlinearity first detraps the particle and after reaching a regime of complete oscillatory motion
it then forces the particle to selftrapping.  As a result nonlinearity first restores the degeneracy that has
been lifted by the presence of more than two sites and subsequently acts in way analogous to the dimer
case, viz.  reaches an equipartition state through a hyperbolic function transition. 

\subsubsection{N=4}
The tetrahedral configuration obtainned for $N=4$ is  rather interesting with 
$\chi_c = \chi_c '=-8$,  $g_2 =g_3 =0$ and the Weierstrass function becoming a simple rational function  
${\cal{P}} (t , 0 , 0 ) = 1/\tau^2 $ leading to the solution
\begin{eqnarray}
p(\tau )=\frac{1-2 {\tau}^2}{1+4 {\tau}^2}
\label{tetramer}
\end{eqnarray}
For long times the solution reaches  the value $-1/2$ while $p_{lin,M}=-1/2$ and
$<p_{lin} > =1/4$ . For positive nonlinearities the localization tendency increases while for negative 
$\chi$ we have first detrapping, complete oscillations for $\chi = -6$ and subsequent asympotic equipartition
at $\chi = -8$ with the transition being now algebraic as seen in Eq. (\ref{tetramer}). Further absolute increase of
$\chi$ produces nonlinear localization. We note that as in the trimer also 
in the present case at $\chi = \chi_c'$ we
have an asymptotic equipartition of the probability on all four sites; in that sence   we could use the
term "selftrapping" meaning however the asymptotic and irreversible onset of equal probability on all sites.

 \begin{widetext}
\begin{figure}[!ht]
\begin{center}
  \includegraphics[height=3.0in,width=3.5in]{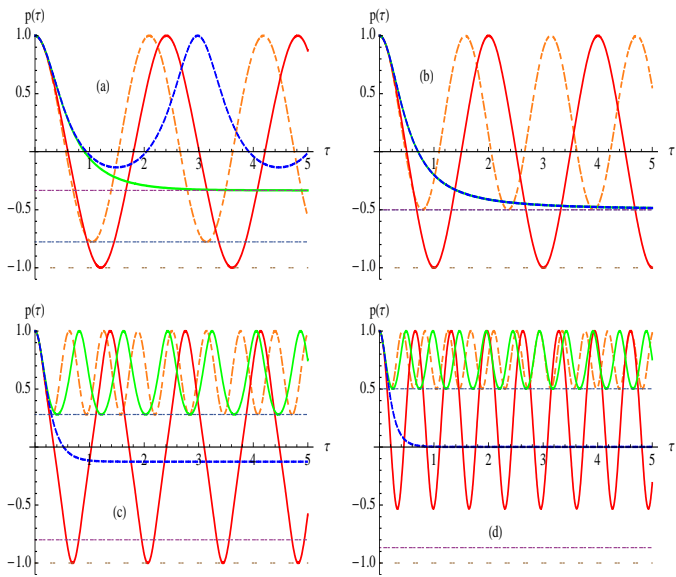}
  \caption{(Color online)  Exact dynamics for fully connected networks: probability $p(\tau )$ of Eq. (\ref{solution})
  as a function of the rescaled time $\tau$. Each subfigure is for a network with different size and specific nonlinearity.
  In all plots the periodic
  continuous red line presents the motion for localized initial condition and for the specific $N$ and $\chi$ while 
  the dashed orange line the solution of the corresponding linear network. The continous green line is the solution at 
  $\chi=\chi_c' \equiv -2N$ while the dashed blue line the solution at $\chi=\chi_c$.  The three horizontal lines parallel to the time axis denote the maximum excursion for the corresponding linear network $p_{lin,M}$ (blue dashed), the equipartition line at $1-1/N$ (
  purple) and the maximum range for the probability $p(\tau )$ at $-1$ (dashed with spaces). In (a) $N=3$ and $\chi =-4$. We note the  restauration of complete oscillations to $-1$ due to nonlinearity and the trigonometric
  evolution (blue curve) that in the present case is distinct from the purely linear motion (dashed orange curve). 
  (b) $N=4$ and $\chi =-6$. The green and blue curves 
  collapse onto a single algebraically decaying solution ($\chi_c = \chi_c' =-8$ that relaxes to the equipartition line,
  while complete periodic motion to the other sites is also observed (red curve reaching $-1$). In this case the maximum linear
  excursion also reaches to the equipartition line. (c) $N=10$ and $\chi = -18$. The solutions for the
  roots at $\chi_c \simeq-19.333$ and $\chi_c' =-20$ have now changed roles with the former (blue line) designating the
  hyperbolic transition (to non-equipartition- equipartition is denoted by the dashed purple line) while the latter (green line) forcing the
  network to pure trigonometric evolution with the same amplitude as for the linear ($\chi =0$) case.  In (d) $N=15$ and 
  $\chi =-15$ we observe similarly the hyperbolic trannsition ( $\chi_c \simeq -27.9772$, blue line) to non-equipartition as
  well as the two trigonometic evolutions for $\chi=0$ (dashed orange line) and $\chi_c'=-20$
  (green line)}
  \end{center}
  \label{Fig1}
\end{figure}

 \end{widetext}
\subsection{$N>4$}
We observed from the previous three cases that in the dimer the two roots at $\chi_c$ and $\chi_c '$
determined two distinct, yet identical in nature, selftrapping transitions. In the trimer $\chi_c'$ was responsible for
the hyperbolic transition, $\chi_c $ for trigonometric evolution,  while the fact that the two roots
were equal in the tetramer turned the (unique) transition to
an algebraic one.    The (negative) roots $\chi_c$ and $\chi_c '$ 
"collide" in the tetramer and we observe that their role changes for $N>4$.  Specifically, we find that
$|\chi_c ' | > | \chi_c |$, with the algebraic transition occurring at $\chi_c$ while at $\chi_c '$ we
have a specific trigonometric evolution.

For negative nonlinearities as we increase in absolute value $\chi$ we have initially detrapping, complete 
oscillatory motion followed by subsequent reduction of the amplitude of the motion that leads to the
hyperbolic transition at $\chi_c$.  This state does not populate equally  at long times  the sites of the $N$-mere as
in the  cases for $N=2,~3,~4$; the asymptotic population of the initially excited site is now larger than that
of the other sites.   Further increase of $|\chi|$ leads to periodic motion  with increasing amplitude reduction.  As
the nonlinearity parameter crosses the value $\chi_c '=-2N$ the dynamics becomes trigonometric while, at the
same time the amplitude of oscillation becomes identical to that of the corresponding linear $N$-mere.  Thus at $\chi_c '$
the nonlinear system behaves identically to the corresponding linear one in what regards the 
site probabilities, nevertheless the evolution is done with different frequency.    Further increase of
nonlinearity leads to more localization.

The value of $\chi_c$ is determined numerically from the solution of the cubic equation
$g(\chi_c )=0$; the values obatained coinside with those found in ref.~\cite{M99}. We observe that as $N$ increases this root separates
from the one at $-2N$ and asymptotically reaches the value $-N$.  Thus in a very large system we expect to
have the hyperbolic selftraping at nonlinearity values of order $-N$ while at $-2N$ the nonlinear system
behaves  as the corresponding linear one with  renormalized frequency.  We also observe that 
the system reaches a regime of   complete oscillations  for $\chi ' = -2(N-1)$ while 
$\chi ' \leq \chi_c' \leq \chi_c |$.
 For  $N  \gtrsim 15$ the equality for $\chi '$ does not hold any more and the system begins to execute incomplete
 oscillations reaching a maximum at $\chi ' \lesssim \chi_c$ . Thus, while for $N < 14$ the particle first
 reaches complete delocalization and at a later stage relaxes hyperbolicaly, for $N \gtrsim 15$ the two phenomena occur
 almost for the same $\chi$-value, first reaching a maximum in the delocalization and subsequently, for
 infinitesimal change in the $\chi$-value, hyperbolic-type evolution. In this last regime the maximum excursing occurs 
 for $\chi$ very close to the hyperbolic transition value $\chi_c$.

The exact expression for the evolution at the trigonometric transition at $\chi_c ' =- 2N$  for $N > 4$ is given
by:

\begin{equation}
p(\tau )= \frac{4-6N +N^2 +2(N-2)  \cos \left( \sqrt{N(N-4)} \tau \right) }
{N\left[ N-2-2\cos \left( \sqrt{N(N-4)} \tau\right)\right] }
\end{equation}

The oscillation extends from $p_{max} =1$ to $p_{min}=(N^2 - 8N + 8)/N^2$; these values are identical to 
the one obtainned for the linear MF limit.  Thus, in the MF limit of the
DNLS equation there are two values of the nonlinearity
parameter, viz. $\chi =0$ and $\chi_c '$ at which an initially 
localized excitation oscillates trigonometrically with the same amplitude,
but with completely differenent frequencies.  In the second case, the finite value of nonlinearity is 
equivalent to a "polaronic" dressing of the excitation leading to a smaller oscillation frequency compared to the
purely linear oscillation at $\chi =0$.

The hyperbolic transition at $\chi_c$ is given, on the other hand, by 

\begin{widetext}
\begin{eqnarray}
p(\tau )=1-
\frac{24(N-1)}{12c\left[ 1+\frac{3}{\sinh^2 [\sqrt{3 c} \tau ]}\right]  +{\chi_c}^2 +2(N-2)\chi_c +N^2 }
\end{eqnarray}
\end{widetext}

where $c$ is the value of the double root of the discriminant of the equation $x^3-g_2 x -g_3 =0$ while the
other root is equal to $-2c$.  

The presence of focusing nonlinearity induces 
complete or partial detrapping to the MF network followed by retrapping for larger $|\chi |$. The exact solution of the nonlinear model for arbitrary $N$ shows thus  very interesting features of detrapping, 
hyperbolic relaxation as well as effective linearization of the nonlinear system.  As the network size $N$ grows,
these interesting fearures appear for increasingly larger values of  the nonlinearity
parameter $|\chi |$, a feature that makes their observation more difficult.
The time evolution for  various networks  is presented in Fig. (1) while the special nonlinearity values 
$\chi_c$ and $\chi_c'$ as a function
of the system size $N$  are portrayed in Fig. (2).
 \begin{figure}[!ht]
\begin{center}
  \includegraphics[height=2.5in,width=3.0in]{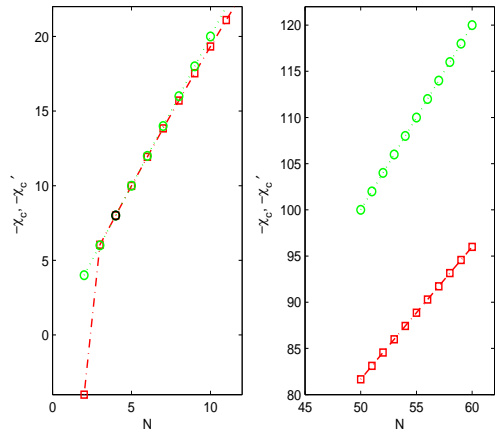}
  \caption{(Color online)  Plot of the values  $-\chi_c$  (red) and $-\chi_c'$ (green) as a function of
  the network size for two regions: Left panel, for $N <12$ and right panel for $50<N<60$. The black circle designates the
  "collision" of the two branches at $N=4$.  For large-$N$ $\chi_c \sim -N$ ~\cite{M99},   while $\chi_c' = -2N$ for any system size.}
  
 \end{center}
  \label{Fig2}
\end{figure}

\section{Conclusions}
The analysis of the MF model for the DNLS equation with a localized initial condition and arbitrary number
of sites $N$ shows that the 
system is rich and has interesting dynamical behaviour.  In the linear case ($\chi =0$) the presence of the long range bonds 
leads to localization that increases with $N$~\cite{PT11}. As focusing nonlinearity is turned on we observe initially a tendency for 
detrapping, i.e. the particle executes oscillations of larger amplitude that can reach complete  or partial ($ N\gtrsim 15$)
 escape from the initially populated site.  

The time dependence of the system is typical of elliptic function dependence, viz. periodic oscillatory motion that includes
an infinite number of frequencies.  This time dependence changes for two specific values of the nonlinearity parameter at
$\chi_c$ and $\chi_c'$ respectively. For the former value the time evolution becomes hyperbolic  (except in the trimer
where the hyperbolic evolution occurs at $\chi_c'$)
and periodic motion ceases at long times.  In the well studied case of $N=2$, the state for this critical nolinearity separates periodic delocalized from localized  motion and we have selftrapping. For $N=3, ~4$, while there is no such separation in the hyperbolic transition, there is asymptotic equipartition of probability for this value.  For $N>4$ the hyperbolic solution at $\chi_c$ simply represents an asymptotic state where motion ceases while
the initially populated site retains most of the probability, a tendency that increases with $N$.  Thus, if by 
"selftrapping" we mean the separation of delocalized to localized motion then, strictly speaking,  it only occurs in the
dimer case. On the other
hand we may extend the term to include the case where asymptotic equipartition of probability takes place; in this case the symmetric trimer and tetramer also qualify, but not  networks with larger number of sites.
 
A new feature
appears at $\chi= \chi_c '$  ($\chi = \chi_c$ in the trimer); 
the elliptic function evolution becomes trigonometric and the infinite set of
frequencies it encompasses collapses to a single one. 
Additionally, for $N>4$, the amplitude of the oscillation becomes identical to the corresponding
one for $\chi =0$. Thus the evolution at $\chi_c'$ is effectively linear but with a new frequency equal to $\sqrt{N(N-4)}$ (for $N>4$)  while in the purely linear case the frequency is $N$~\cite{PT11}.  If we think
that the nonlinear term stems from the coupling with additional degrees of freedom ~\cite{HT99} then the evolution becomes 
more sluggish since the particle is dressed and carries with it the additional degrees of freedom.  It is remarkable that
such a  dressing leading to frequency renormalization may occur and still the evolution be trigonometric  with the
same amplitude as in the purely linear case.  In the large-$N$ limit the two frequencies approach one the other.

As the size of the system increases we observe two distinct features, one at $\chi_c \sim -N$ where 
time-hyperbolic behaviour takes place while at $\chi_c'= -2N$ where we have an effective linearization of the motion.
Both features appear  after the initial detraping and retraping has occurred as a function of $|\chi |$.  The 
onset of nonlinear localization is typically associated with discrete breather solutions that are localized in space 
and time  periodic. The  case for $\chi = \chi_c '$ found  here would then correspond to a "linear breather", i.e. a
localized nonlinear solution with exact trigonometric evolution.  It would be interesting to find out if a similar situation can occur in other discrete nonlinear systems as well.

\end{document}